\documentclass[twocolumn,prb]{revtex4}% Physical Review B
\usepackage{graphicx}% complex graphics
\usepackage{bm}% bold math
\usepackage{amssymb} %math symbols
\begin{document}

\title{Microscopic theory of surface-enhanced Raman scattering
 in noble-metal nanoparticles}

\author{Vitaliy N. Pustovit$^{1,2}$ and Tigran V. Shahbazyan$^{1}$}

\affiliation{$^{1}$Department of Physics and Computational Center for Molecular
  Structure and Interactions, Jackson State University, Jackson, MS
  39217, USA \\
  $^{2}$Laboratory of Surface Physics, Institute of Surface
  Chemistry, Kyiv 03164, Ukraine} 

%\date{July 13, 2004} 

\begin{abstract}
We present a microscopic model for surface-enhanced Raman scattering
(SERS) from molecules adsorbed on small noble-metal nanoparticles. 
In the absence of direct overlap of molecular orbitals and electronic
states in the metal, the main enhancement source is the strong
electric field of the surface plasmon resonance in a nanoparticle 
acting on a molecule near the surface. In small particles, the
electromagnetic enhancement is strongly modified by quantum-size
effects. We show that, in nanometer-sized particles, SERS magnitude is 
determined by a competition between several quantum-size effects such
as the Landau damping of surface plasmon resonance and reduced
screening near the nanoparticle surface. 
Using time-dependent local density approximation, we calculate
spatial distribution of local fields near the surface and 
enhancement factor for different nanoparticles sizes.  
\end{abstract}


%% \begin{abstract}
%% A microscopic approach to surface-enhanced Raman scattering (SERS)from
%% molecules adsorbed on noble-metal nanoparticles is developed. For
%% nanoparticle sizes smaller than 10 nm, the classical electromagnetic
%% enhancement mechanism is modified by quantum-size effects. Using
%% time-dependent local field approximation, we perform systematic
%% analysis of SERS in nanometer-sized Ag nanoparticles. We find that, in
%% small nanoparticles, Raman crossection enhancement is governed by the
%% interplay between Landau damping of the surface plasmon and interband
%% screening in the nanoparticle surface layer.  
%% \end{abstract}

\maketitle
% \begin{twocolumn}

\section{Introduction}
Surface-enhanced Raman scattering (SERS) has been one of the
highlights of optical spectroscopy in metal nanostructures during past
25 years. \cite{schatz-review02} Recent interest in SERS
%\cite{fleischman-cpl74,vanduyne-jec77}
stems from the discovery of extremely strong single-molecule SERS in
silver nanoparticle aggregates, \cite{nie-sci97,kneipp-prl97} as well
as from nanoparticle-based applications such as, e.g., biosensors
\cite{mirkin-science02} that rely on sensitivity of SERS to small
concentrations of target molecules.  The main mechanism of SERS has
long been known as electromagnetic (EM) enhancement
\cite{schatz-review02,moskovits-rmp85,kerker-ao80,gersten-jcp81} of
dipole moment of a molecule by the strong local field of surface
plasmon (SP) resonance in a nanoparticle.  EM mechanism is especially
effective when a cluster of nanoparticles is concentrated in a small
region (``hot spot'').
\cite{kneipp-cr99,brus-jpcb00,moskovits-tap02,rothberg-pnas04} A
combined effect of SP local fields from different particles acting on
a molecule trapped in a gap can result in a giant (up to $10^{14}$)
enhancement of the Raman scattering crossection.
\cite{stockman-prb96,markel-prb96,kall-prl99,kall-prb00,corni-jcp02,stockman-prl03}
Other mechanisms contributing to SERS can involve electron tunneling
between a molecule and a nanoparticle. \cite{otto-jpcm92}

The conventional description of EM enhancement is based on classical
Mie scattering theory. \cite{kerker-ao80,gersten-jcp81}  The dipole
moment of a molecule at distance ${\bf r}_0$ form a particle center is
enhanced by a factor $\sim\alpha_p(\omega)/r_0^3$, where $\alpha_p =
R^3 \bigl[(\epsilon-1)/(\epsilon+2)\bigr]$ is the particle polarizability,
$R$ is its radius, and $\epsilon(\omega)$ is metal dielectric
function. The far-field of molecular dipole, radiating at
Stokes-shifted frequency $\omega_s$, is, in turn, comprised of direct
and Mie-scattered fields. The latter contributes another factor
$\sim\alpha_p(\omega_s)/r_0^3$, so that the total field enhancement is
$\sim\alpha_p(\omega)\alpha_p(\omega_s)/r_0^6$ and Raman crossection is
proportional to $|\alpha_p|^4/r_0^{12}$. At frequencies close to the
SP pole in $\alpha_p$, this enhancement can reach $\sim 10^6$.
Note that, within classical description, the dependence of SERS on
nanoparticle size, coming from geometrical factor in $\alpha$, is weak
if the molecule is sufficiently close to nanoparticle surface.

The classical approach is valid for relatively large nanoparticles,
where the effect of confining potential on electronic states is
negligible. For nanoparticle sizes $R \lesssim 5$ nm, the lifetime of
SP is reduced due to the Landau damping by 
single-particle excitations accompanied by momentum transfer to the
surface.\cite{kubo-jpsj66} This results in a broadening of SP
resonance peak by the amount of level spacing at the Fermi energy,
$\gamma_s\sim v_F/R$ ($v_F$ is the Fermi velocity), 
and in the corresponding decrease of the SP field amplitude. For not
very small nanoparticles, the effect of Landau damping on SERS can be
treated semiclassically \cite{schatz-review02}
by incorporating the quantum-size correction $\gamma_s$ in the Drude
dielectric function of metal. Recent resonance fluorescence
measurements on small gold particles,
\cite{feldmann-prl02,feldmann-nl05,nie-jacs02} however, indicate strong 
deviations of the plasmon-induced enhancement from that predicted by
semiclassical models. \cite{gersten-jcp81}
For clusters with electron number $N<100$, SP lifetime is reduced to
several fs and SERS is diminished. Note that some enhancement of Raman
signal due to single-particle resonances remains even in small
clusters containing several atoms. \cite{dickson-prl05}

In this paper, we study SERS for small nanoparticles several
nanometers in diameter, i.e., in the intermediate regime between classical
particles and small clusters. Our chief observation is that, in this
crossover regime, SERS magnitude is determined by interplay between several
competing quantum-size effects, including the aforementioned SP Landau
damping as well as the modification of electron screening in the surface
region. The latter produces an opposite trend towards a 
{\em relative increase} of SERS when the molecule is 
located in a close proximity to the metal surface. The underlying
mechanism is related to different effects that the confining potential has
on {\em d}-band and {\em sp}-band electron states.  Namely, the deviation of
potential well from the rectangular shape gives rise to a larger 
{\em effective} radius for the higher-energy {\em sp}-electrons.
\cite{persson-prb85} This effect is further amplified by the 
{\em spillover} of {\em sp}-band electron density, due to tunneling
into the barrier, as contrasted to the essentially step-like density
profile of localized {\em d}-electrons.  As a result, in a 
{\em surface layer} of thickness $1-2$ {\AA}, the 
{\em d}-electron population is diminished and, hence, the interband
(i.e. due to {\em d--sp} transitions) polarizability is strongly reduced.
\cite{liebsch-prb93,liebsch-prb95}
The resulting {\em reduction of screening} in the
surface region leads to a greater strength of electron-electron
interactions in the {\em sp}-band that was observed, e.g., in a faster,
as compared to bulk metal, electron relaxation in Ag nanoparticles
measured using ultrafast pump-probe spectroscopy. \cite{voisin-prl00}
It is, therefore, natural to expect  
that such underscreening should lead to additional enhancement of
local field {\em outside} of the nanoparticle. Note, however, that
here the effect of spillover is twofold: while it leads to an increase
of volume fraction of underscreened region and hence to stronger local
field, especially in small nanoparticles, it can also, by itself, have an
opposite effect by smearing out the otherwise sharp classical
boundary. \cite{dignam-jcp92} Thus, in small nanoparticles, the
enhancement magnitude is determined 
by a delicate interplay of competing quantum-size effects, and must,
therefore, be described within a consistent microscopic approach. Such
an approach, based on time-dependent local density approximation (TDLDA),
\cite{ekardt-prb85} is developed in this paper.
  
The paper is organized as follows. In Section \ref{sec:polar}, we
derive the general expression for polarizability of molecule-nanoparticle
system. In Section \ref{sec:factor}, we derive a self-consistent system
for local potential that determines the enhancement factor. In
Section \ref{sec:num} we present the results of our numerical
calculations. Section \ref{sec:conc} concludes the paper.

\section{Polarizability of molecule-nanoparticle system}
\label{sec:polar}

\begin{figure}
%  \begin{center}
  \centering
  \includegraphics[width=3.0in]{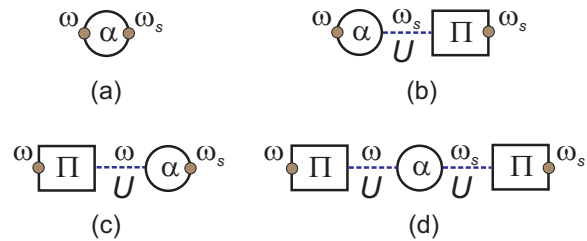}
%  \end{center}
  \caption{\label{fig:diagram} Nonradiative processes contributing to Raman
  scattering from a molecule-nanoparticle system.
 }
  \end{figure}

We start with formulating SERS in terms of quantum transitions in the
interacting molecule-nanoparticle system.  We assume that the molecule
is located at distance $r_0$ from the nanoparticle center and that the
overall system size is much smaller than radiation wavelength so the
retardation effects can be ignored. \cite{kreibig-book}  In the
absence of direct electron tunneling, \cite{otto-jpcm92} the
interactions within excited molecule-nanoparticle system are caused by
{\em nonradiative transitions} accompanied by energy transfer between
a molecule and a nanoparticle, similar to Forster transfer in
two-molecule systems. \cite{lakowicz-book} Namely, an electron-hole
pair can nonradiatively recombine by transferring its energy to SP
(and vice versa) via dynamically-screened Coulomb interaction.
\cite{shahbazyan-prl98} Feynman diagrams of processes contributing
to polarizability of molecule-nanoparticle system, $\tilde{\alpha}$,
are shown in Fig.\ \ref{fig:diagram}: (a) incident photon with energy
$\omega$ is absorbed by the molecule and reemitted with Stokes-shifted
energy $\omega_s$; (b) after absorbing a photon, excited molecule
nonradiatively recombines transferring its energy to SP in the
nanoparticle, which emits a photon; (c) SP, excited by incident light,
transfers its energy to the molecule, which emits a photon; and (d)
after energy transfer from SP to molecule, the latter transfers the
energy back to SP, which emits a photon. Correspondingly, the system
polarizability is (in operator form) 
\begin{eqnarray}
\label{diagram}
\tilde{\alpha}=\alpha+\alpha U\Pi + \Pi U\alpha + \Pi U \alpha U \Pi ,
\end{eqnarray}
where $\alpha$ is molecular polarizability, $\Pi$ is
density-density response function of a nanoparticle in medium, and $U$
is the Coulomb potential.  The Raman polarizability is obtained by
calculating the matrix element of $\tilde{\alpha}$ between incoming
and outgoing photon states with energies $\omega$ and $\omega_s$,
respectively. The interaction of the molecule with the nanoparticle
involves 
matrix elements $\langle e|\phi({\bf r}) |g \rangle$, where $|g
\rangle$ and $|e \rangle$ stand for the molecule ground and excited
electronic bands, respectively, and
\begin{eqnarray}
\label{response}
\phi(\omega,{\bf r})=
\int d{\bf r}_1d{\bf r}_2 U({\bf r}-{\bf r}_1) 
\Pi(\omega,{\bf r}_1, {\bf r}_2)\phi_0 ({\bf r}_2)
\end{eqnarray}
is the nanoparticle response to external photon potential, 
$\phi_0 ({\bf r})$.  Since the length scale of $\phi({\bf r})$ is much
larger than the molecule size, we have $\langle e|\phi({\bf
r})|g\rangle \approx \bm{\mu}\nabla\phi({\bf r}_0)$, where $\bm{\mu}$
is the dipole matrix element of corresponding molecular transition.
\cite{long-book} The averaging over random orientations of $\bm{\mu}$
can be accounted by assuming isotropic Raman polarizability tensor
$\alpha$. The nanoparticle contribution then factors out,
$\tilde{\alpha}=\alpha M$, with
\begin{eqnarray}
\label{factor}
M=1+\frac{1}{E_0^2}\Bigl[ {\bf E}_0
\cdot 
\nabla \phi(\omega,{\bf r}_0) + 
{\bf E}_0
\cdot
\nabla \phi(\omega_s,{\bf r}_0) 
\qquad
\nonumber\\
+ \nabla\phi(\omega,{\bf r}_0)
\cdot 
\nabla \phi(\omega_s,{\bf r}_0)\Bigr],
\end{eqnarray}
where ${\bf E}_0$ is the electric field in the absence of
nanoparticle. For incident field, ${\bf E}_i$, polarized along the
$z$-axis, $\phi_0=-eE_i r \cos\theta$, we have ${\bf E}_0=
{\bf E}_i/\epsilon_m$, where $\epsilon_m$ is the dielectric constant
of medium. 
% 
%  %%%%%%%%%%%%%%%%%%%%%%%%%%%%%%%%%%
%  %%%%%%%%%%%%%%%%%%%%%%%%%%%%%%%%
% 

\section{Enhancement factor}
\label{sec:factor}

To evaluate the local potential $\phi (\omega,{\bf r})$
within TDLDA approach, we present it in the form
\begin{eqnarray}
\label{poisson}
\phi (\omega,{\bf r})
%% =\int d{\bf r}_1d{\bf r}_2 U({\bf r}-{\bf r}_1) 
%% \Pi(\omega,{\bf r}_1, {\bf r}_2)\phi_0 ({\bf r}_2)
%% \nonumber
=e^2 \int d^3r' \,
\frac{\delta n(\omega,{\bf r}')}{|{\bf r}-{\bf r}'|},
\end{eqnarray}
where the induced density,
\begin{equation}
\label{density-decompose}
\delta n({\bf r})=\int d{\bf r}'\Pi({\bf r}, {\bf r}')
\phi_0({\bf r}')
=
\delta n_s({\bf r})+\delta n_d({\bf r}) +\delta n_m({\bf r}),
\end{equation} 
contains contributions from {\em sp}-electrons, {\em d}-electrons, and
surrounding medium, respectively (hereafter we suppress frequency
dependence).

We adopt the two-region model that combines a quantum-mechanical
description for {\em sp}-band electrons and phenomenological treatment
{\em d}-electrons with bulk-like ground-state density $n_d$ in the
region confined by $R_d<R$. \cite{liebsch-prb93} This model has been
used for calculations of polarizabilities of small Ag nanoparticles
and clusters, \cite{kresin-prb95,lerme-prl98} but it remains reliable
for relatively large electron numbers, $N>1000$.  The induced density
of {\em sp}-band electrons is determined from TDLDA equation
\begin{eqnarray}
\label{tdlda}
\delta n_s({\bf r}) = \int d^3 r' P_s ({\bf r}, {\bf r}') \Bigl[\Phi({\bf r}')
+ V'_x[n(r')]\delta n_s ({\bf r}')\Bigr],
\end{eqnarray}
where $\Phi=\phi_0+\phi$ is the full potential, $P_s ({\bf r}, {\bf
r}')$ is the polarization operator for noninteracting 
{\em sp}-electrons, $V'_x[n(r')]$ is the (functional) derivative of the
exchange-correlation potential and $n(r)$ is the ground-state electron
density. The latter is obtained in a standard way by solving Kohn-Sham
equations.  To close the system, we need to express the full potential
$\Phi({\bf r})$ via $\delta n_s({\bf r})$. This is accomplished be
relating $\delta n_d({\bf r})$ and $\delta n_m({\bf r})$ back to
$\Phi({\bf r})$ as 
\begin{eqnarray}
\label{back}
e^2 \delta n_d({\bf r})= \nabla
\bigl[\chi_d(r)\nabla \Phi({\bf r})\bigr],
\nonumber\\
e^2 \delta n_m({\bf r})= 
\nabla \bigl[\chi_m(r)\nabla \Phi({\bf r})\bigr],
 \end{eqnarray}
where
$\chi_d(r)=\bigl[(\epsilon_d-1)/4\pi\bigr] \,\theta(R_d-r)$ is the interband
susceptibility with the step function enforcing the boundary
conditions and, correspondingly, $\chi_m(r)=\bigl[(\epsilon_m-1)/4\pi\bigr]
\,\theta(r-R)$ is the susceptibility of surrounding medium. The
derivation is given in Appendix. The final result is conveniently
expressed in terms of expansion of $\Phi({\bf r})$ and 
$\delta n({\bf r})$ in spherical harmonics. For {\em non-resonant} Raman
scattering, only dipole ($L=1$) terms contribute to the local
field. The final expressions for dipole component $\Phi(r)$ can be
presented as a decomposition (see the Appendix) 
 \begin{equation}
 \label{phi}
\Phi=\frac{1}{\epsilon(r)} \bigl[\phi_0(r)+\delta\phi_d(r)
+\delta \phi_s(r)\bigr], 
% \Phi=\varphi_0+\delta \varphi_0+ \delta \varphi_s,
 \end{equation}
where 
%$\varphi_0=\phi_0/\epsilon(r)=-eE_ir/\epsilon(r)$,
$\phi_0(r)=-eE_ir$,
\begin{eqnarray}
\label{delta-phi-0}
%% \delta \phi_d(r) = 
%% \beta\left(\frac{r}{R}\right) \phi_0(R)\,
%% \frac{\lambda_m (1-a^3\lambda_d)}{1-2a^3\lambda_d\lambda_m}
%% \qquad
%% \nonumber\\
%% - \beta\left(\frac{r}{R_d}\right) \phi_0(R_d)\, 
%%  \frac{\lambda_d (1-2\lambda_m)}{1-2a^3\lambda_d\lambda_m},
%
\delta \phi_d(r) = 
%\frac{1}{\epsilon(r)} \Bigl[ 
\beta(r/R) \phi_0(R)\, \lambda_m(1-a^3\lambda_d)/\eta 
\qquad 
\nonumber\\
-\beta(r/R_d) \phi_0(R_d)\, \lambda_d (1-2\lambda_m)/\eta
% \Bigr],
\end{eqnarray}
and 
\begin{eqnarray}
\label{delta-phi-s}
\delta \phi_s(r)= \int dr' r'^2 K(r,r')\delta n_s(r').
\end{eqnarray}
Here $\epsilon(r)= (\epsilon_d$, 1, $\epsilon _m$) for $r$ in the
intervals [$(0,R)$, $(R_d,R)$, $(R,\infty)$], 
respectively, and 
\begin{eqnarray}
\label{lambda}
\lambda_d=\frac{\epsilon_d-1}{\epsilon_d+2}, ~~
\lambda_m=\frac{\epsilon_m-1}{2\epsilon_m+1}, ~~
%% \nonumber\\
\eta=1-2a^3\lambda_d\lambda_m,
\end{eqnarray}
with $a=R_d/R$. The kernel $K(r,r')$, relating the induced potential
and density of {\em sp}-electrons, is given by
%\begin{widetext}
%
\begin{eqnarray}
\label{A}
K(r,r') = u(r,r')
\qquad \qquad \qquad \qquad \qquad \qquad \qquad
\nonumber\\
-
\beta(r/R_d)\Bigl[u(R_d,r') 
- 
2a \lambda_m u(R,r')\Bigr]\lambda_d/\eta
\nonumber\\
+
 \beta(r/R)\Bigl[u(R,r') - a^2 \lambda_d
 u(R_d,r')\Bigr]\lambda_m/\eta,
%\biggr],
\end{eqnarray}
where $u(r,r')=4\pi r_</3r_>^2$ is the dipole term of Coulomb
potential expansion and $\beta(x)=x^{-2}\theta(x-1)-2x\theta(1-x)$.
With decomposition (\ref{phi}), the TDLDA equation (\ref{tdlda})
takes the form 
\begin{eqnarray}
\label{tdlda2}
%&&
\delta n_s(r) = \int d r' r'^2 P_s (r,r') \frac{1}{\epsilon(r')}
\Bigl[\phi_0(r')+\delta\phi_d(r')\Bigr] 
\qquad
\nonumber\\
+
\int d r'  r'^2 P_s (r,r') \frac{1}{\epsilon(r')}
\Biggl[\int d r'' r''^2  K(r',r'')\delta n_s (r'')
\nonumber\\
+ V'_x(r')\delta n_s (r')\Biggr],
\qquad 
%
%% \delta n_s(r) =\int d r' r'^2 P_s (r,r') 
%% \Bigl[\varphi_0(r')+\delta\varphi_0(r')\Bigr]
%% \qquad \qquad
%%  \nonumber\\
%% %&&
%% \qquad
%% +
%% \int d r'  r'^2 P_s (r,r') \biggl[\int d r'' r''^2  K(r',r'')\delta n_s (r'')
%% \nonumber\\
%% + V'_x(r')\delta n_s (r')\biggr].
%% \qquad
\end{eqnarray}
%
%\end{widetext}
Note that  $\Phi(r)$ is continuous at $r=R_d,R$. 

Equations\ (\ref{phi})--(\ref{tdlda2}) determine self-consistently the
spatial distribution of local potential near small noble-metal
nanoparticles. Here $\delta\phi_d(r)$ is the induced potential due
to {\em d}-electrons and surrounding medium. Their effect on the 
{\em sp}-electron potential, $\delta\phi_s(r)$, is encoded in the
kernel $K(r,r')$.  For $\epsilon_d=\epsilon_m=1$, we have
$K(r,r')=u(r,r')$ and $\delta\phi_d(r)=0$, recovering the case of
simple metal particles in vacuum.  

If the molecule is not too close to
the surface ($d\gtrsim 1${\AA}), i.e., there is no significant overlap
between molecular orbitals and electronic states, then 
Eqs.\ (\ref{poisson}) and (\ref{phi})--(\ref{tdlda2}) yield 
\begin{equation}
\label{delta-phi}
\delta \Phi(r_0)\equiv 
\frac{1}{\epsilon(r_0)}\,\bigl[\delta\phi_d(r_0)+\delta\phi_s(r_0)\bigr]
=\frac{eE_i}{\epsilon_m r_0^2}\, \alpha_p,
\end{equation}
where  $\alpha_p(\omega)$ 
%% %
%% \begin{equation}
%% \label{alpha-p}
%% \alpha_p
%% = \frac{4\pi}{3eE_i} \int dr r^3\delta n(r)
%% \end{equation}
%% %
is the nanoparticle polarizability. The expression for
$\alpha_p$ is given in Appendix. Field enhancement coefficient $M$, 
Eq.\ (\ref{factor}), then takes the form
\begin{eqnarray}
\label{enhance-field}
M=1+(1+\cos^2\theta_0)\,
%[g(\omega)+g(\omega_s)]
\frac{\alpha_p(\omega)+\alpha_p(\omega_s)}{r_0^3}
\nonumber\\
+(1+3\cos^2\theta_0)\,
\frac{\alpha_p(\omega)\, \alpha_p(\omega_s)}{r_0^6}\, .
% g(\omega)g(\omega_s),
\end{eqnarray}
%
%with $g=\alpha_p/r_0^3$.
In this case,
enhancement retains the same functional dependence on particle
polarizability as in classical theory;\cite{kerker-ao80} however,
$\alpha_p$ is now determined microscopically.

%%%%%%%%%%%%%%%%%%%%%%%%%%%%%%%%%%%%%%%%%%%%%%%%%%%%%%%%%%%%
\section{Numerical results and discussion}
\label{sec:num}

\begin{figure}[t]
 \begin{center}
% \centering
\includegraphics[width=2.6in]{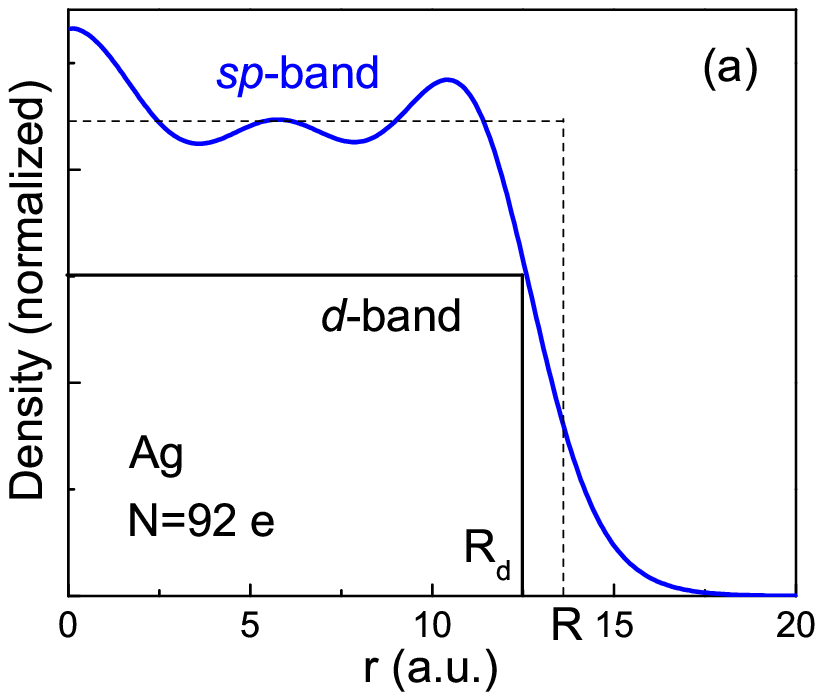}
\includegraphics[width=2.6in]{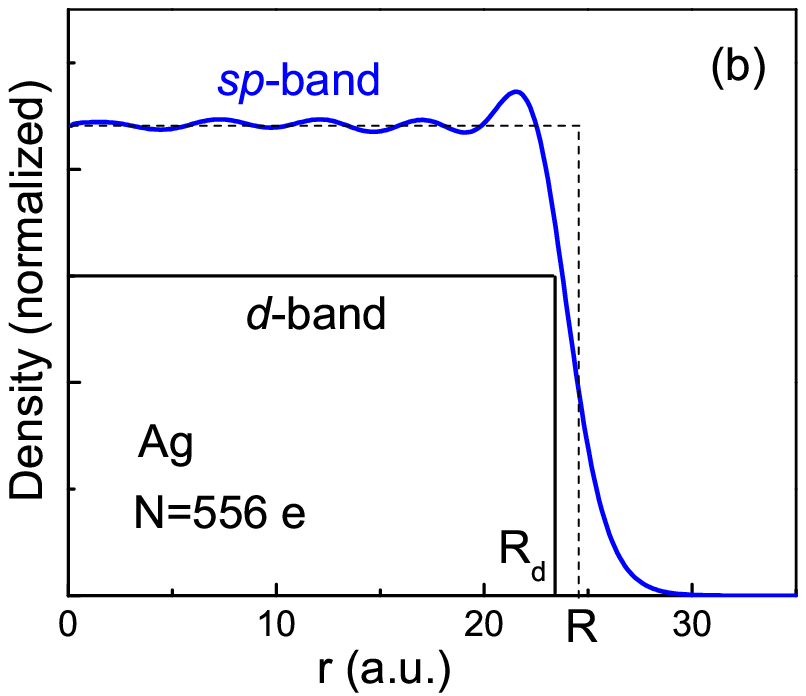}
\includegraphics[width=2.6in]{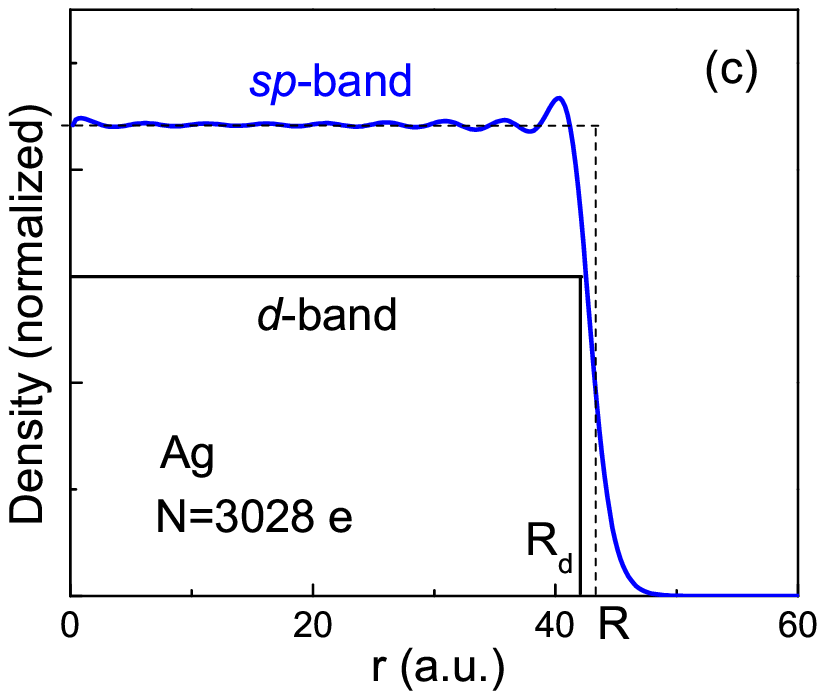}
 \end{center}
 \caption{\label{fig:density} Calculated ground state density $n(r)$ for Ag
   nanoparticles is shown for closed-shell electron numbers $N=92$
   (a), $N=556$ (b), and $N=3028$ (c).
}
 \end{figure}

%{\noindent \em Numerical results} --- 
Below we present our results for SERS enhancement factor $|M|^2$
for Ag nanoparticles in a medium with dielectric constant
$\epsilon_m=1.5$.  Calculations
were carried for number of electrons ranging from $N=92$ to $N=3028$,
corresponding to particle diameters in the range $D\approx 1.4 - 4.5$
nm. To ensure spherical symmetry, only closed-shell``magic numbers''
were used. \cite{koch-prb96} For such sizes, the Ag band-structure
remains intact. The ground state energy spectrum and wave-functions
were obtained by solving the Kohn-Sham equations for jellium model
\cite{ekardt-prb85} with the Gunnarsson-Lundqvist exchange-correlation
potential; \cite{lundqvist-prb77} the interaction strength was
appropriately modified to account for static {\em d}-band
screening. The ground state density of {\em sp}-band electrons,
$n(r)$, exhibits characteristic Friedel oscillations, while the
spatial extent of spillover is $\simeq 2$ a.u. 
(see Fig.\ \ref{fig:density}). 
%% This value is somewhat 
%% smaller than for nanoparticles in vacuum due to the additional
%% screening of Coulomb  interactions by surrounding medium. 
%%%%%%%%%%%%%%%%%%%%%%%%%%%%%%%%%%%
%%%%%%%%%%%%%%

These results were used as input in the numerical solution 
of TDLDA system\ (\ref{phi})--(\ref{tdlda2}). The molecule was located
along the 
$z$-axis ($\theta_0=0$) at a distance $d=5$ a.u. from effective
boundary with radius $R=r_sN^{1/3}$, where $r_s=(4\pi n/3)^{-1/3}$
($r_s=3.0$ a.u. for Ag), ensuring no direct overlap with the
nanoparticle.  In the calculation of optical response, the experimental
data for $\epsilon_d(\omega)$ in Ag was used \cite{palik-book}.
We also assumed that the Stokes shift is much smaller than
$\gamma_s\sim v_F/R$ ($\gamma_s\simeq 0.5$ eV for $D=3.0$ nm) and
ignored the difference between $\omega$ and $\omega_s$. 
%
%%%%%%%%%%%%%%%%%%%%%%%%%%%%%%%%%%%
 \begin{figure}
 \begin{center}
% \centering
\includegraphics[width=2.7in]{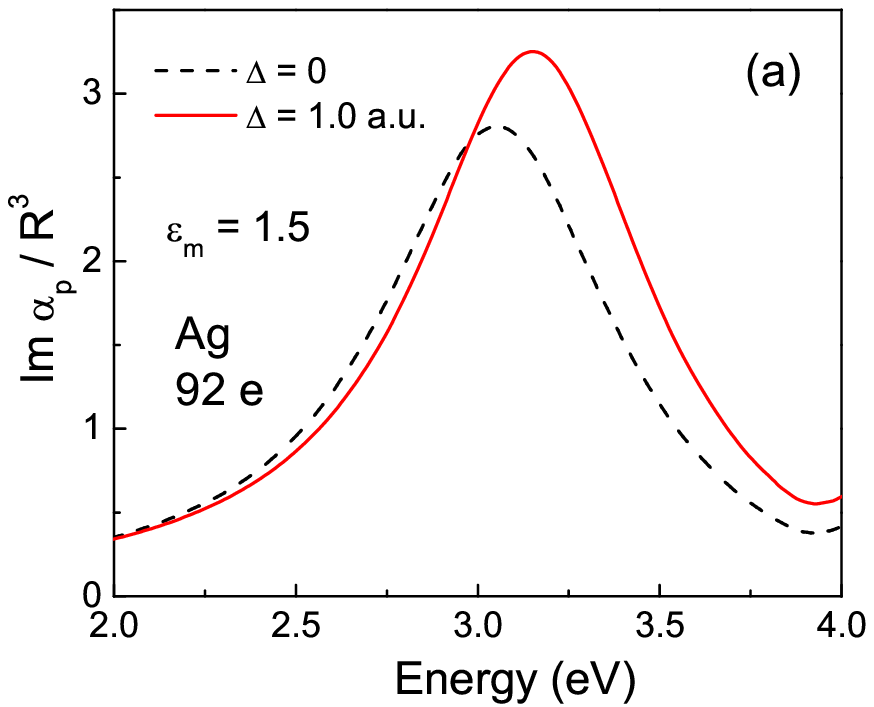}
\includegraphics[width=2.7in]{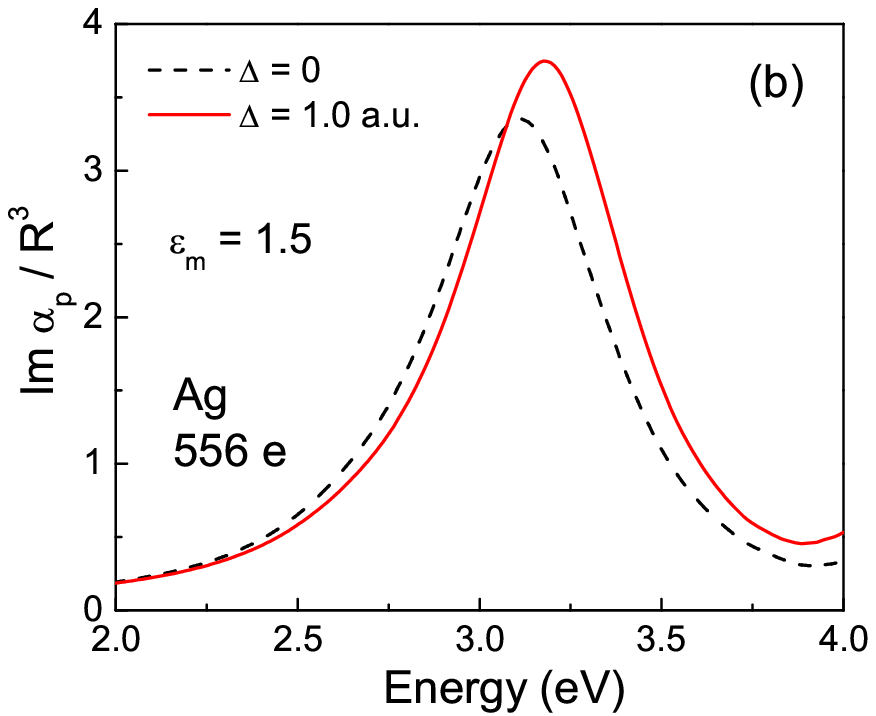}
\includegraphics[width=2.7in]{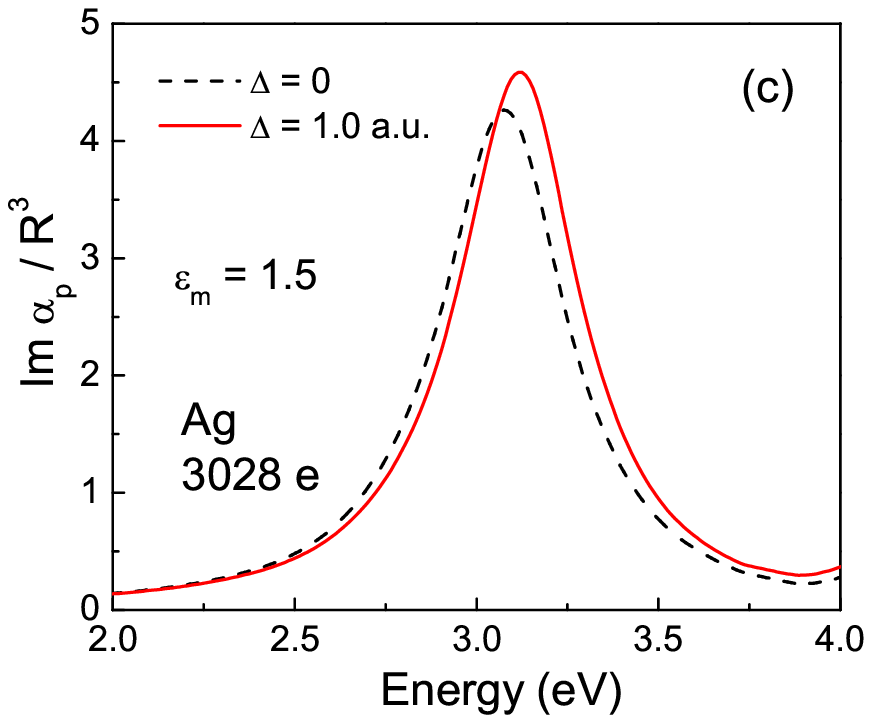}
 \end{center}
 \caption{\label{fig:absorp} Calculated absorption spectra for Ag
   nanoparticles with $\Delta=0$ (dashed line) and $\Delta=1.0$ a.u. (solid
   line) is shown for electron numbers $N=92$
   (a), $N=556$ (b), and $N=3028$ (c).
}
 \end{figure}
%%%%%%%%%%%%%%

The calculated absorption spectra for different nanoparticle sizes are
shown in Fig.~\ref{fig:absorp}. For medium dielectric constant
$\epsilon_m=1.5$, the position of SP resonance at $\simeq 3.2$ eV is
well below of the interband transition onset at $\simeq 4.0$ eV. As
expected, with increasing size the peak width is reduced due to a
weaker Landau damping of SP. In order to illustrate the role of
interband screening, we also show the results of calculations with
$\Delta=0$. Note that decreasing the surface layer thickness by 1.0
a.u. only somewhat reduces the underscreening; the main effect still
comes from the larger spatial extent (about 2 a.u.) of 
{\em sp}-band electron spillover. The effect of underscreened surface
region on the absorption is two-fold. The large contrast
ratio of $\epsilon_d$ in the bulk and surface regions 
[$\epsilon_d(\omega_{sp})\approx 5$] leads to a lower, as compared
to bulk, {\em average} value of interband dielectric function in the
relevant frequency region. As a result, the peak position for $\Delta=1.0$ 
a.u. is slightly blueshifted with respect to that for 
$\Delta=0$ ($\delta\omega_{sp} \sim 0.05$ eV). At the same time, for
$\Delta=1.0$ a.u., the peak {\em amplitude} is larger although the
resonance width stays unchanged. The stronger absorption for
$\Delta=1.0$ a.u. is caused by a weaker screening of the SP electric field,
that determines the peak oscillator strength, in the surface region.

\begin{figure}
% \begin{center}
 \centering 
 \includegraphics[width=2.6in]{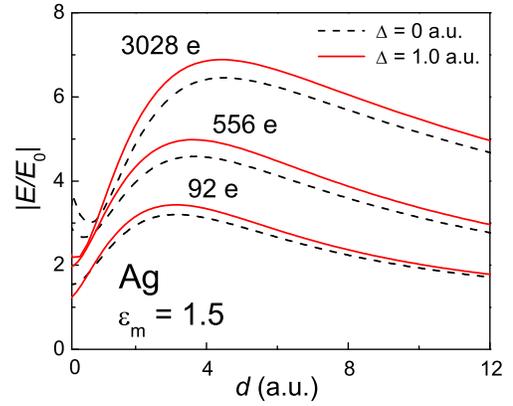}
% \end{center}
 \caption{\label{fig:field} Calculated local field for several
   nanoparticles sizes vs. distance from classical boundary for
   $\Delta=0$ and $\Delta=1.0$ a.u.}
 \end{figure}

The calculated local field, $E$, at resonance frequency is plotted in
Fig.\ \ref{fig:field} vs. molecule-nanoparticle distance,
$d=r_0-R$. The gradual rise of field  magnitude on the length scale of
electron spillover replaces the discontinuity (for
$\epsilon_d,\epsilon_m\neq 1$) of classical field 
at the sharp boundary.  It can be seen that while at $\Delta=1.0$
a.u. the field amplitude is larger than for $\Delta=0$, the difference
decreases for larger $d$. Correspondingly, the effect of interband
screening  on Raman signal enhancement, $|M|^2\propto |E|^4$, is
substantial only when the molecule is located close to the nanoparticle.
%
%%%%%%%%%%%%%%%%%%%%%%%%%%%%%%%%%%%
%%%%%%%%%%%%%%%%%%%%%%%%%%%%%%%%%

\begin{figure}[t]
% \begin{center}
 \centering
\includegraphics[width=3.2in]{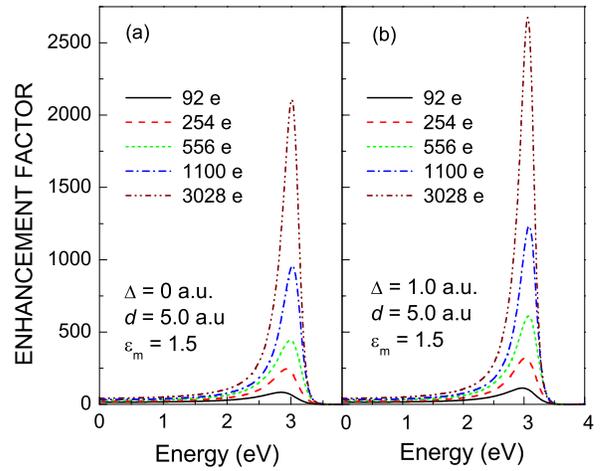}
% \end{center}
 \caption{\label{fig:enhancement} Calculated enhancement factor
   $|M|^2$ for different
   nanoparticle sizes is shown with $\Delta=0$ (a) and $\Delta=1.0$ a.u. (b).
}
 \end{figure}

\begin{figure}
% \begin{center}
\centering
\includegraphics[width=2.8in]{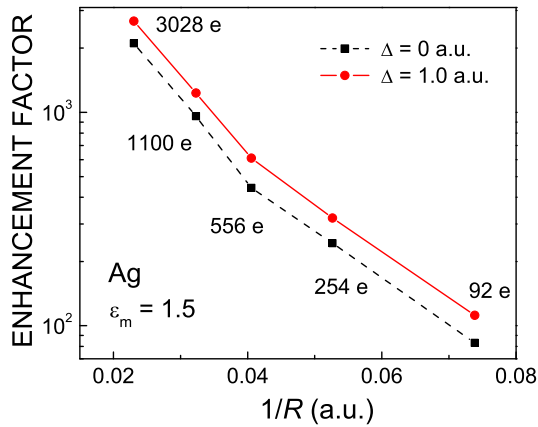}
\caption{\label{fig:size} Calculated enhancement factor $|M|^2$ at
  resonance frequency vs. nanoparticle size is shown with $\Delta=0$
  and $\Delta=1.0$ a.u. 
} 
\end{figure}

In Fig.\ \ref{fig:enhancement}, we plot the enhancement factor $|M|^2$
as a function of incident light energy for different nanoparticle
sizes. Note that the asymmetric shape of the enhancement peak, as
compared to the absorption peak, is because the former is determined
by the absolute value of polarization, rather than its imaginary
part. The general tendency is a decrease of SERS for smaller
nanoparticles with the enhancement factor reaching only $|M|^2\sim 100$
for the smallest nanoparticle size, $D\approx 1.4$ nm. This decrease
is mainly related to the Landau damping of SP: at resonance energy, we
have $E\propto 1/\gamma \sim R/v_F$ yielding $|M|^2\propto R^4$. The
comparison of results for $\Delta=0$ $\Delta=1.0$  a.u. shown in 
Figs.\ \ref{fig:enhancement}(a) and \ref{fig:enhancement}(b),
respectively, shows that 
reducing the thickness of surface layer even by 1.0 a.u. substantially
affects the enhancement. The role of interband screening is  most
visible in the dependence of SERS on nanoparticle size plotted in
Fig. \ref{fig:size} at resonance frequency for 
nanoparticle diameters in the range 1.4--4.5 nm. With decreasing size,
the enhancement factor drops by an order of magnitude, while overall
enhancement is reduced for $\Delta=0$. 
Importantly, the latter effect is more pronounced for smaller
nanoparticles; for $\Delta=1.0$ a.u., the enhancement factor is larger
by 25\% for $N=3028$ but by 35\% for $N=92$ than for $\Delta=0$
indicating a more important role of screening in smaller nanoparticles
due to larger surface-to-volume ratio. Thus, the proper account of
screening gives a {\em slower} (as compared to semiclassical models)
decrease of the enhancement as nanoparticles become smaller. Note,
finally, that by keeping $\Delta$ constant for different nanoparticle
sizes, we somewhat underestimated the screening contribution to SERS;
indeed, for smaller nanoparticles, underscreened layer is thicker due
to stronger deviations of the confining potential from rectangular shape.
%
%%%%%%%%%%%%%%%%%%%%%%%%%%%%%%%%%%%
%%%%%%%%%%%%%%%%%%%%%%%%%%%%%%%%%

%
%%%%%%%%%%%%%%%%%%%%%%%%%%%%%%%%%
% %%%%%%%%%%%%%%%%%%%%%%%%%%%%%%%%
%

\section{Conclusions}
\label{sec:conc}

We developed a microscopic model for surface-enhanced Raman scattering
in noble-metal nanoparticles. Our approach incorporates, in a
unified manner, all relevant quantum-size effects that determine the
magnitude of Raman signal enhancement for nanometer-sized particles. 
While the Landau damping of surface plasmons leads to a 
general decrease of the enhancement for small particles, this trend is
partially offset by the reduction of interband screening in the metal
boundary region. The additional enhancement of local field is
substantial only in a close proximity to the metal surface, and it is
more pronounced for smaller nanoparticles where the electron density
profile deviates strongly from the classical shape

As a final remark, we considered here a {\em nonresonant} Raman
scattering, i.e. the case when the  excitation energy of a molecule is much
larger than the SP energy $\omega_{sp}$. In this case, due to small
single-molecule polarizability, only linear (in molecule) response
needs to be considered and SERS is dominated by a {\em single} back
and forth energy transfer within molecule-nanoparticle system [see
Fig.\ \ref{fig:diagram}(d)]. In the resonant case, {\em multiple} energy
transfer processes give rise to a nonradiative width of the molecular
levels and, in general, to a reduction of the
enhancement.\cite{kall-prl04} For small molecule-surface separations,
these nonradiative processes are dominated by surface-enhanced
electron-hole pair generation in the metal. Even in simple metals,
these processes are enhanced due to a reduced {\em intraband}
(Thomas-Fermi) screening near the
boundary. \cite{persson-prb82,stockman-prb04} The issue of
nonradiative decay for noble-metal nanoparticles, where  
{\em interband} screening effects become important, will be addressed
in a future publication.  

\acknowledgements
This work was supported by NSF under Grants No. DMR-0305557 and
NUE-0407108, by NIH under Grant No. 5 SO6 GM008047-31, and by ARL
under Grant No. DAAD19-01-2-0014.

\appendix*
\section{Derivation of local potential}

The full self-consistent potential 
$\Phi({\bf r})=\phi ({\bf r})+\phi_0({\bf r})$ sutisfies Poisson equation
\begin{eqnarray}
\label{poisson-full}
\Phi (\omega,{\bf r})
=\phi_0({\bf r})+ e^2 \int d^3r' \,
\frac{\delta n(\omega,{\bf r}')}{|{\bf r}-{\bf r}'|},
\end{eqnarray}
where the induced density si comprised of {\em sp}-band, {\em d}-band
and medium contriburions, 
$\delta n({\bf r})=
\delta n_s({\bf r})+\delta n_d({\bf r})+\delta n_m({\bf r})$. 
Using Eq.\ (\ref{back}) for $\delta n_m$ and $\delta n_d$ and
integrating by parts, Eq.\ (\ref{poisson-full}) takes the form
\begin{eqnarray}
\label{poisson1}
\epsilon(r)\Phi({\bf r}) 
= 
\phi_0({\bf r}) + e^2\int d^3r' 
\frac{\delta n_s({\bf r}')}{|{\bf r}-{\bf r}'|}
\qquad \qquad ~~~
\nonumber\\
+ 
\frac{\epsilon_d-1}{4\pi} \int d^3r' 
\nabla'\frac{1}{|{\bf r}-{\bf r}'|}
\cdot \nabla'\theta(R_d-r) \Phi({\bf r}')
\nonumber\\
+
\frac{\epsilon_m-1}{4\pi} \int d^3r'
\nabla'\frac{1}{|{\bf r}-{\bf r}'|}\cdot \nabla'\theta(r-R) \Phi({\bf r}'),
\end{eqnarray}
where $\epsilon(r)= (\epsilon_d$, 1, $\epsilon _m$) for $r$ in the
intervals [$(0,R)$, $(R_d,R)$, $(R,\infty)$], respectively. Since the
source term has the form  
$\phi_0({\bf r})=\phi_0(r)\cos\theta=-e E_ir\cos\theta$, we expand
$\Phi$ and $\delta n_s$ in terms of spherical harmonics and, keeping
only the dipole term ($L=1$), obtain
\begin{eqnarray}
\label{poisson2}
\epsilon(r)\Phi(r) 
=
\phi_0(r) + e^2 \int dr'r'^2 u(r,r')\delta u_s(r')
\nonumber\\
- \frac{\epsilon_d-1}{4\pi} R_d^2 \,
\frac{\partial u(r,R_d)}{\partial {R_d}} \, \Phi(R_d)
\nonumber\\
+
\frac{\epsilon_m-1}{4\pi} R^2  \,
\frac{\partial u(r,R)}{\partial {R}} \, \Phi(R),
\end{eqnarray}
where 
\begin{equation}
\label{coulomb} 
u(r,r') =\frac{4\pi}{3}\biggl[\frac{r'}{r^2} \,
\theta(r-r') +\frac{r}{r'^2} \, \theta(r'-r)\biggr]
\end{equation}
is the dipole term of the radial component of the Coulomb
potential. The above equation can be simplified to
\begin{eqnarray}
\label{poisson-nano}
\epsilon(r) \Phi(r) = \bar{\phi}(r)
- \frac{\epsilon_d-1}{3} \beta(r/R_d)\Phi(R_d)
\nonumber\\
+ \frac{\epsilon_m-1}{3} \beta(r/R) \Phi(R),
\end{eqnarray}
where 
$\beta(r/R)=\bigl(3R^2/4\pi\bigr)\bigl[\partial u(r,R)/\partial R\bigr]$
is given by
\begin{equation}
\label{beta}
\beta(x) = x^{-2}\, \theta (x-1) -2x \theta(1-x),
\end{equation}
and we introduced a shorthand notation
\begin{eqnarray}
\label{phi-s}
\bar{\phi}(r) =  \phi_0(r)+ e^2 \int dr'r'^2 u(r,r')\delta n_s(r').
\end{eqnarray}
The boundary values of $\Phi$ can be obtained by matching $\Phi(r)$
at $r= R_d, R$, yielding
\begin{eqnarray}
\label{boundary}
(\epsilon_d+2)\Phi(R_d)+2a(\epsilon_m-1)\Phi(R)=3\bar{\phi}(R_d),
\nonumber\\
(\epsilon_d-1)a^2\Phi(R_d)+(2\epsilon_m+1)\Phi(R)=3\bar{\phi}(R),
\end{eqnarray}
where $a=R_d/R$. 
Substituting $\Phi(R_d)$ and $\Phi(R)$ back into
Eq.\ (\ref{poisson-nano}), we arrive at
\begin{eqnarray}
\label{poisson-nano1}
\epsilon(r) \Phi(r) = \bar{\phi}(r)
- \beta(r/R_d) \, \frac{\lambda_d}{\eta}\,
\Bigl[\bar{\phi}(R_d)-2a \lambda_m \bar{\phi}(R)\Bigr] 
\nonumber\\
+
\beta(r/R) \, \frac{\lambda_m}{\eta}\,
\Bigl[\bar{\phi}(R)-a^2\lambda_d\bar{\phi}(R_d)\Bigr],
\qquad
\end{eqnarray}
where the cofficients $\lambda$ are given by 
Eq.\ (\ref{lambda}). Separating out $\delta n_s$-dependent
contribution, we arrive at Eq.\ (\ref{phi}).

%%%%%%%%%%%%%%%%%%%%%%%%%%%%%%%%%%%%%%%%%%%%%%%%%%%%%%%%%%%%%
The expression for nanoparticle polarizability,
$\alpha_p(\omega)$, can be obtained from the large-$r$ asymptotics of
induced potential, Eq.\ (\ref{delta-phi}).
From Eqs. (\ref{delta-phi-0})--(\ref{A}), we find for $r\gg R$
\begin{equation}
\label{delta-phi-a}
\delta \phi_d(r) = \frac{eE_i}{ r^2}\, \alpha_d,
~~~
\delta \phi_s(r) = \frac{eE_i}{r^2}\, \alpha_s,
\end{equation}
with
\begin{eqnarray}
%\hspace{-2in}
\alpha_d
=
R^3\,
%% \left[1 - \frac{(1+\lambda_m)(1-a^3\lambda_d)}{1-2a^3\lambda_d\lambda_m}
%% \right]
\frac{(1-\lambda_m)a^3\lambda_d-\lambda_m}{1-2a^3\lambda_d\lambda_m},
%\hspace{3in}
%% \nonumber\\
%% \alpha_s(\omega_s)
%% &=&
\qquad \qquad \qquad ~~~
\nonumber\\
\alpha_s=
\frac{4\pi}{3eE_i}
\Biggl[\int_0^{\infty}dr'r'^3\delta n_s(r')
\qquad \qquad \qquad ~~~
\nonumber\\
-
%\Bigl[1 -\frac{(1+\lambda_m)(1-\lambda_d)}{1-2a^3\lambda_d\lambda_m} \Bigr]
\frac{\lambda_d-\lambda_m +\lambda_d(1-2a^3)}{1-2a^3\lambda_d\lambda_m}\,
\int_0^R dr'r'^3 \delta n_s(r')
\nonumber\\
-
%% \Bigl[1
%% -\frac{(1+\lambda_m)(1-a^3\lambda_d)}{1-2a^3\lambda_d\lambda_m}\Bigr]
\frac{(1-\lambda_m)a^3\lambda_d-\lambda_m}{1-2a^3\lambda_d\lambda_m}\,
R^3\int_R^{\infty}dr'\delta n_s(r')
\nonumber\\
+
\frac{(1+\lambda_m)\lambda_d}{1-2a^3\lambda_d\lambda_m}\,
\int_{R_d}^{R}dr'(r'^3-R_d^3)\delta n_s(r')
\Biggr].
%\nonumber
\end{eqnarray}
%
%% %
%% \begin{eqnarray}
%% \label{alpha}
%% \alpha_d(\omega_s)
%% =
%% R^3\Bigl[1 - (1+\lambda_m)(1-a^3\lambda_d)/\eta\Bigr],
%% \qquad \qquad 
%% \nonumber\\
%% \alpha_s(\omega_s)
%% =
%% \frac{4\pi}{3eE_i}
%% \Biggl[\int_0^{\infty}dr'r'^3\delta n_s(r')
%% \qquad \qquad \qquad \qquad 
%% \nonumber\\
%% -
%% \Bigl[1 -(1+\lambda_m)(1-\lambda_d)/\eta\Bigr]
%% \int_0^R dr'r'^3 \delta n_s(r')
%% \nonumber\\
%% -
%% \Bigl[1 -(1+\lambda_m)(1-a^3\lambda_d)/\eta\Bigr]
%% R^3\int_R^{\infty}dr'\delta n_s(r')
%% \nonumber\\
%% +
%% \Bigl[(1+\lambda_m)\lambda_d/\eta\Bigr]
%% \int_{R_d}^{R}dr'(r'^3-R_d^3)\delta n_s(r')
%% \Biggr].
%% \qquad 
%% \end{eqnarray}
%% %
The nanoparticle polarizability is $\alpha_p=\alpha_d+\alpha_s$. Note
that for frequencies below interband absorption onset, the {\em d}-band/medium
contribution $\alpha_d$ is real.

\end{document}